\documentclass[aps,prb,superscriptaddress,twocolumn,10pt,longbibliography]{revtex4-2}
\usepackage{amsmath, amssymb}
\usepackage{braket}
\usepackage{color}
\usepackage{graphicx}
\usepackage{ulem}
\usepackage{siunitx}

\begin{document}
\date{\today}
\title{First-principles study on tunnel magnetoresistance effect with Cr-doped RuO$_{2}$ electrode}
\author{Katsuhiro Tanaka}
\affiliation{Department of Physics, University of Tokyo, Hongo, Bunkyo-ku, Tokyo 113-0033, Japan}
\author{Takuya Nomoto}
\affiliation{Department of Physics, Tokyo Metropolitan University, Hachioji, Tokyo 192-0397, Japan}
\affiliation{Research Center for Advanced Science and Technology, University of Tokyo, Komaba, Meguro-ku, Tokyo 153-8904, Japan}
\author{Ryotaro Arita}
\affiliation{Research Center for Advanced Science and Technology, University of Tokyo, Komaba, Meguro-ku, Tokyo 153-8904, Japan}
\affiliation{Center for Emergent Matter Science, RIKEN, Wako, Saitama 351-0198, Japan}
\begin{abstract}
We investigate the functionality of the $\mathrm{Cr}$-doped $\mathrm{RuO_{2}}$ as an electrode of the magnetic tunnel junction (MTJ),
motivated by the recent experiment showing that $\mathrm{Cr}$-doping into the rutile-type $\mathrm{RuO_{2}}$ will be an effective tool to control its antiferromagnetic order and the resultant magnetotransport phenomena easily.
We perform first-principles calculation of the tunnel magnetoresistance (TMR) effect in the MTJ based on the $\mathrm{Cr}$-doped $\mathrm{RuO_{2}}$ electrodes.
We find that a finite TMR effect appears in the MTJ originating from the momentum-dependent spin splitting in the electrodes, 
which suggests that $\mathrm{RuO_{2}}$ with Cr-doping will work as the electrode of the MTJ.
We also show that this TMR effect can be qualitatively captured using the local density of states inside the tunnel barrier.
\end{abstract}
\maketitle
\section{Introduction}
The tunnel magnetoresistance (TMR) effect is a spin-dependent transport phenomenon observed in a magnetic tunnel junction (MTJ), 
a multilayered system consisting of insulating thin films sandwiched by magnetic electrodes~\cite{Julliere1975_PhysLettA_54A_225}.
The tunnel resistance can differ between when the magnetic moments of two magnetic electrodes align parallelly or antiparallelly.
\par
The TMR effect has long been discussed with ferromagnetic electrodes~\cite{Julliere1975_PhysLettA_54A_225,Maekawa1982_IEEETranMagn_18_707,Slonczewski1989_PhysRevB_39_6995,Miyazaki1995_JMagnMagnMater_139_L231,Moodera1995_PhysRevLett_74_3273,Mathon1997_PhysRevB_56_11810,Butler2001_PhysRevB_63_054416,Mathon2001_PhysRevB_63_220403,Parkin2004_NatMater_3_862,Yuasa2004_NatMater_3_868,Ikeda2008_ApplPhysLett_93_082508,Ikeda2010_NatMater_9_721,Worledge2011_ApplPhysLett_98_022501,Scheike2023_ApplPhysLett_122_112404,Tsymbal2003_JPhysCondensMatter_15_R109,Zhang2003_JPhysCondensMatter_15_R1603,Ito2006_JMagnSocJpn_30_1,Yuasa2007_JPhysD_40_R337,Butler2008_SciTechnolAdvMater_9_014106}
since it has been believed that macroscopic spin polarization is essential to generate a TMR effect.
However, recent studies have revealed that MTJs with antiferromagnets can also exhibit a finite TMR effect~\cite{Chen2023_Nature_613_490,Merodio2014_ApplPhysLett_105_122403,Stamenova2017_PhysRevB_95_060403,Zelezny2017_PhysRevLett_119_187204,Jia2020_SciChinnaPhys_63_297512,Shao2021_NatCommun_12_7061,Smejkal2022_PhysRevX_12_011028,Dong2022_PhysRevLett_128_197201,Qin2023_Nature_613_485,Tanaka2023_PhysRevB_107_214442,Cui2023_PhysRevB_108_024410,Gurung2023_arXiv_2306.03026,Jia2023_PhysRevB_108_104406,Jiang2023_PhysRevB_108_174439,Chi2024_PhysRevApplied_21_034038,Samanta2023_arXiv_2310.02139,Shi2023_arXiv_2311.13828,Zhu2024_JMagnMagnMater_597_172036,Shao2023_arXiv_2312.13507}. 
Particularly, the antiferromagnets breaking the time-reversal symmetry macroscopically are promising materials to show the TMR effect;
when the magnetic order of an antiferromagnet breaks the time-reversal symmetry macroscopically,
namely, when a magnetically ordered state does not return to the original state after the time-reversal operation and succeeding translation or inversion operations, 
a finite spin splitting in the momentum space is generated~\cite{Noda2016_PhysChemChemPhys_18_13294,Ahn2019_PhysRevB_99_184432,Naka2019_NatCommun_10_4305,Smejkal2020_SciAdv_6_eaaz8809,Smejkal2022_PhysRevX_12_031042,Smejkal2022_PhysRevX_12_040501}.
When such an antiferromagnet is used for the electrode of the MTJ, 
the momentum-dependent spin splitting contributes to generating the difference in the transmission of two configurations.
The MTJs using such antiferromagnets have been discussed theoretically~\cite{Shao2021_NatCommun_12_7061,Smejkal2022_PhysRevX_12_011028,Dong2022_PhysRevLett_128_197201,Chen2023_Nature_613_490,Qin2023_Nature_613_485,Cui2023_PhysRevB_108_024410,Gurung2023_arXiv_2306.03026,Jiang2023_PhysRevB_108_174439,Chi2024_PhysRevApplied_21_034038,Samanta2023_arXiv_2310.02139,Shi2023_arXiv_2311.13828}, 
and a finite TMR effect has been actually observed in experiments~\cite{Chen2023_Nature_613_490,Qin2023_Nature_613_485,Shi2023_arXiv_2311.13828}.
\par
A typical material whose antiferromagnetic structure breaks the time-reversal symmetry macroscopically is the rutile-type $\mathrm{RuO_{2}}$.
Recently, $\mathrm{RuO_{2}}$ has found to possess the collinear antiferromagnetic structure ~\cite{Berlijn2017_PhysRevLett_118_077201,Zhu2019_PhysRevLett_122_017202}.
Triggered by this finding, 
its spin splitting nature in the momentum space~\cite{Ahn2019_PhysRevB_99_184432,Fedchenko2024_SciAdv_10_eadj4883} and the magnetotransport phenomena
such as the anomalous Hall effect~\cite{Smejkal2020_SciAdv_6_eaaz8809,Feng2022_NatElectro_5_735} or the N\'{e}el spin current~\cite{Shao2023_PhysRevLett_130_216702} have been discussed. 
As for the TMR effect, 
the TMR ratio has been theoretically calculated in the $\mathrm{RuO_{2}}(001)/\mathrm{TiO_{2}}/\mathrm{RuO_{2}}$-~\cite{Shao2021_NatCommun_12_7061}, 
$\mathrm{RuO_{2}}(110)/\mathrm{TiO_{2}}/\mathrm{RuO_{2}}$-~\cite{Jiang2023_PhysRevB_108_174439},
and $\mathrm{RuO_{2}}/\mathrm{TiO_{2}}/\mathrm{CrO_{2}}$-MTJs~\cite{Chi2024_PhysRevApplied_21_034038,Samanta2023_arXiv_2310.02139} as well as in the argument based on the properties of the electrode~\cite{Smejkal2022_PhysRevX_12_011028}.
\par
Furthermore, a recent experiment investigated the physical properties of the Cr-doped $\mathrm{RuO_{2}}$~\cite{Wang2023_NatCommun_14_8240},
where it was shown that the $\mathrm{Cr}$-doped $\mathrm{RuO_{2}}$ exhibits a distinct anomalous Hall effect in a small or zero magnetic field.
This experiment has suggested that doping Cr is a possible way to easily manipulate the magnetic order and magnetotransport properties of $\mathrm{RuO_{2}}$ systems.
\par
In this paper, we discuss the functionalities of the $\mathrm{Cr}$-doped $\mathrm{RuO_{2}}$ as an electrode of the MTJ.
First we perform first-principles calculation of the $\mathrm{Cr}$-doped $\mathrm{RuO_{2}}$ and show that $\mathrm{Cr}$-doped $\mathrm{RuO_{2}}$ has the spin splitting depending on the momentum.
Then, we perform first-principles calculation of the TMR effect with 
$\mathrm{Ru}_{1-x}\mathrm{Cr}_{x}\mathrm{O}_{2}/\mathrm{TiO_{2}}/\mathrm{Ru}_{1-x}\mathrm{Cr}_{x}\mathrm{O}_{2}$ MTJ
and find that the $\mathrm{Cr}$-doped $\mathrm{RuO_{2}}$ shows a finite TMR effect owing to that spin splitting structure.
\par
This paper will also have a practical significance in designing MTJs with compensated magnets.
We have recently proposed a method to evaluate the TMR effect qualitatively with the local density of states (LDOS) inside the tunnel barrier in the model calculation~\cite{Tanaka2023_PhysRevB_107_214442}.
Namely, by using this method, the calculation of the transmission, 
which usually demands a high computational cost in first-principles calculations, could be skipped.
However, that proposal is based on the idealized lattice model.
It is not trivial whether we can apply the estimation in terms of the local density of states for more realistic MTJ systems, 
particularly the TMR effect with compensated magnets.
We verify the applicability of the estimation method in terms of the LDOS.
\section{System and Method}
\subsection{Density functional theory calculations of Cr-doped $\mathrm{RuO_{2}}$ and tunnel magnetoresistance effect with Cr-doped $\mathrm{RuO_{2}}$}
\begin{figure}[tbh]
	\centering
	\includegraphics[width=86mm]{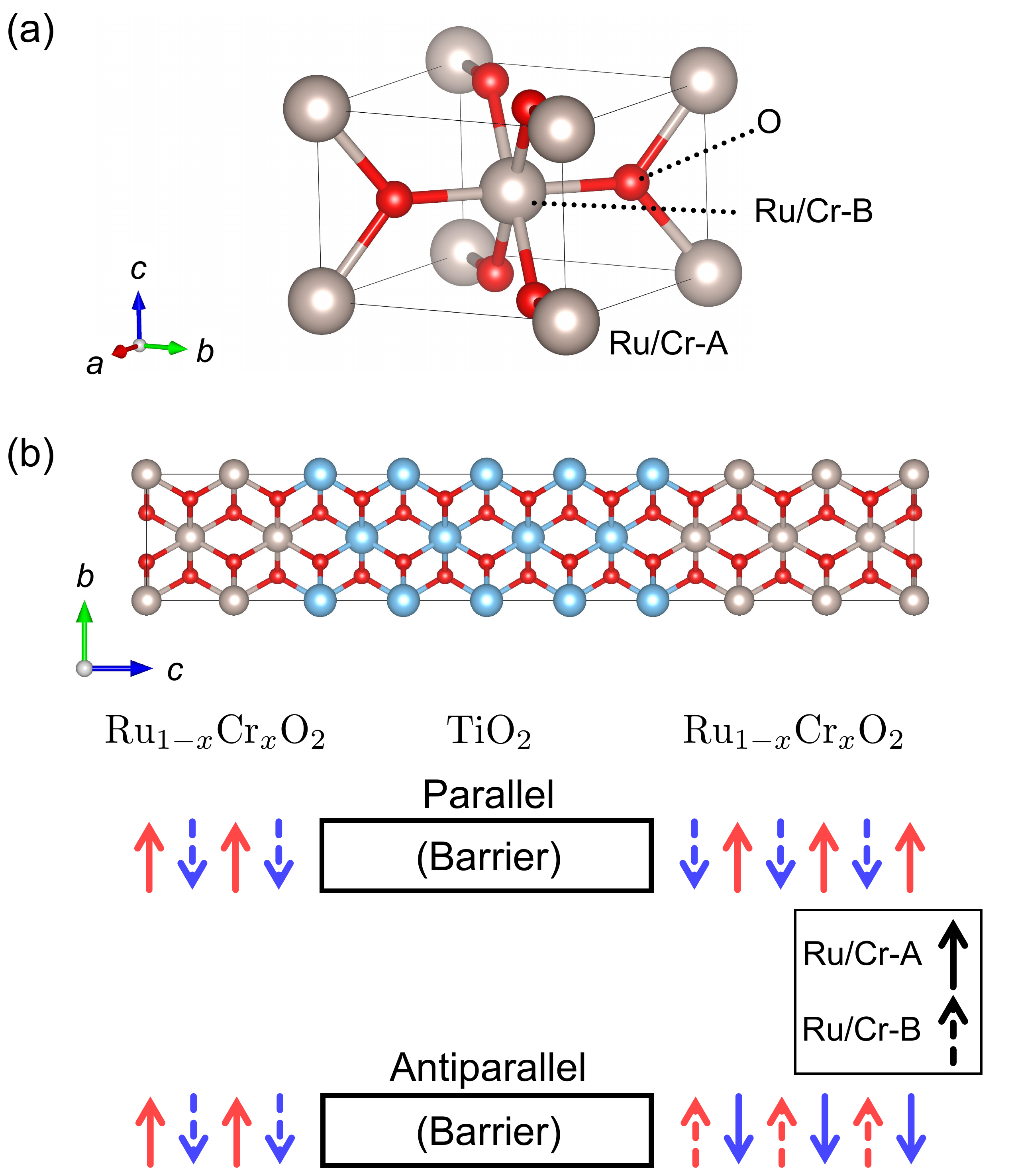}
	\caption{%
		(a) Crystal structure of $\mathrm{Ru}_{1-x}\mathrm{Cr}_{x}\mathrm{O}_{2}$.
		(b) Crystal structure of the $\mathrm{Ru}_{1-x}\mathrm{Cr}_{x}\mathrm{O}_{2}/\mathrm{TiO_{2}}/\mathrm{Ru}_{1-x}\mathrm{Cr}_{x}\mathrm{O}_{2}$ magnetic tunnel junction and schematics of the alignments of the magnetic moments for the parallel and antiparallel configurations.
		Arrows and arrows with broken lines represent the magnetic moments of the Ru/Cr-A and Ru/Cr-B sites, respectively.
	}
	\label{fig:MTJstructure}
\end{figure}
We use the rutile-type $\mathrm{RuO_{2}}$ whose $\mathrm{Ru}$-sites are partially substituted for $\mathrm{Cr}$, $\mathrm{Ru}_{1-x}\mathrm{Cr}_{x}\mathrm{O}_{2}$,
as the electrode of the MTJ.
$\mathrm{RuO_{2}}$ has the tetragonal crystal structure (space group: $P4_{2}/mnm$) with two inequivalent $\mathrm{Ru}$-sites, $\mathrm{Ru}$-A and $\mathrm{Ru}$-B, in a unit cell (Fig.~\ref{fig:MTJstructure}(a)).
We use the $a$- and $c$-axis lattice constants of $\mathrm{RuO_{2}}$ as $a_{\mathrm{RuO_{2}}} = 4.4919$~{\AA} and $c_{\mathrm{RuO_{2}}} = 3.1066$~{\AA}, respectively.
The position of oxygen atom is $x = 0.3058$ for $\mathrm{RuO_{2}}$.
\par
As a barrier layer of the MTJ, we use the rutile-type $\mathrm{TiO_{2}}$.
Namely, we calculate the TMR effect in the $\mathrm{Ru}_{1-x}\mathrm{Cr}_{x}\mathrm{O}_{2}/\mathrm{TiO_{2}}/\mathrm{Ru}_{1-x}\mathrm{Cr}_{x}\mathrm{O}_{2}$ MTJ. 
Here, $\mathrm{Ru}_{1-x}\mathrm{Cr}_{x}\mathrm{O}_{2}$ and $\mathrm{TiO_{2}}$ are stacked along (001) direction (Fig.~\ref{fig:MTJstructure}(b)).
For $\mathrm{TiO_{2}}$, the in-plane lattice constant is matched to $a_{\mathrm{RuO_{2}}}$,
and $c$-axis length is $c_{\mathrm{TiO_{2}}} = 2.9589$~{\AA}.
The interface between $\mathrm{RuO_{2}}$ and $\mathrm{TiO_{2}}$ is given by the average of $c_{\mathrm{RuO_{2}}}$ and $c_{\mathrm{TiO_{2}}}$.
The position of oxygen atom for $\mathrm{TiO_{2}}$ is $x = 0.3057$.
\par
The TMR effect is calculated based on the scattering theory approach~\cite{Choi1999_PhysRevB_59_2267} with the Landauer--B\"{u}ttiker formula~\cite{Landauer1957_IBMJResDev_1_3,Landauer1970_PhilMag_21_863,Buttiker1986_PhysRevLett_57_1761,Buttiker1988_IBMJResDevelop_32_317}.
First, we separate the whole MTJ into three parts: the left and right leads, and the scattering region.
Here, the left and right leads are $\mathrm{Ru}_{1-x}\mathrm{Cr}_{x}\mathrm{O}_{2}$,
and the scattering region is nine monolayers (MLs) of $\mathrm{TiO_{2}}$ with four MLs and five MLs of $\mathrm{Ru}_{1-x}\mathrm{Cr}_{x}\mathrm{O}_{2}$ on its left and right sides, respectively.
We calculate the electronic structure of each of these three parts.
Then, we construct the MTJ attaching these three parts and calculate the transmission.
\par
To obtain the electronic structure, we perform the density functional theory (DFT) calculation~\cite{Hohenberg1964_PhysRev_136_B864,Kohn1965_PhysRev_140_A1133} using the \textsc{Quantum ESPRESSO (QE)} package~\cite{Giannozzi2009_JPhysCondensMatter_21_395502,Giannozzi2017_JPhysCondensMatter_46_465901}.
We use the norm-conserved pseudopotential obtained from \textsc{PseudoDojo}~\cite{Hamann2013_PhysRevB_88_085117,vanSetten2018_ComptPhysCommun_226_39}.
The exchange correlation is taken in by the Perdew--Berke--Ernzerhof type generalized gradient approximation~\cite{Perdew1996_PhysRevLett_77_3865}.
The energy cutoff for the wave-function is 110~Ry, and that for the charge density is 440~Ry.
We take $15 \times 15 \times 20$ and $15 \times 15 \times 1$ $\boldsymbol{k}$-points for the self-consistent field (scf) calculation of the lead and the scattering region, respectively.
We do not consider the effect of spin-orbit coupling or the additional Coulomb interaction, $+U$.
We use the virtual crystal approximation (VCA) to substitute Ru with Cr.
For the antiparallel configuration, 
we attach the copy of the scattering region with its magnetic structure inverted and calculate the electronic structure of the doubled scattering region to deal with the electronic and magnetic structures at the boundary between the leads and the scattering region properly.
When we calculate the transmission, the doubled supercell is cut in half and restored to the original scattering region.
Here, the parallel and antiparallel configurations are defined by focusing on the same sublattices;
the MTJ has the parallel (antiparallel) configurations when $\mathrm{Ru}/\mathrm{Cr}$-A sites of the left and right electrodes have the magnetic moments aligned parallelly (antiparallelly) (Fig.~\ref{fig:MTJstructure}(b)).
\par
For the calculation of the transmission, we use the \textsc{pwcond} codes contained in the \textsc{QE} package~\cite{Smogunov2004_PhysRevB_70_045417,DalCorso2005_PhysRevB_71_115106,DalCorso2006_PhysRevB_74_045429}.
Following the Landauer--B\"{u}ttiker formula, the total conductance, $G$, 
is given by the total transmission, $T_{\text{tot}}$, 
as $G = (e^{2}/h) T_{\text{tot}}$ with the elementary charge $e$ and the Planck constant $h$.
The total transmission $T_{\text{tot}}$ is calculated by summing up the transmission at each in-plane $\boldsymbol{k}_{\parallel} = (k_{x}, k_{y})$ point perpendicular to the conducting path with spin-$\sigma$, 
$T_{\sigma}(\boldsymbol{k}_{\parallel})$, as, 
\begin{align}
	T_{\text{tot}} 
& = \sum_{\sigma = \uparrow, \downarrow}\dfrac{1}{N_{\boldsymbol{k}_{\parallel}}}\sum_{\boldsymbol{k}_{\parallel}} T_{\sigma}(\boldsymbol{k}_{\parallel}).
\end{align}
Here, the $z$-direction which is the conducting direction is taken along the $c$-axis of the MTJ (see also Fig.~\ref{fig:MTJstructure}(b)), 
and $N_{\boldsymbol{k}_{\parallel}}$ is the number of $\boldsymbol{k}_{\parallel}$-point in the transmission calculation.
We take $N_{\boldsymbol{k}_{\parallel}} = 251 \times 251$.
\subsection{Evaluation of tunnel magnetoresistance effect with local density of states}
Here, we briefly review the method which is used to qualitatively estimate the TMR effect with LDOS based on Ref.~\cite{Tanaka2023_PhysRevB_107_214442}.
Using the conventional Jullire's picture,
the transmission is approximated by the spin polarization, 
or equivalently, the total DOS, 
of the two magnetic metals used for the electrodes~\cite{Julliere1975_PhysLettA_54A_225,Maekawa1982_IEEETranMagn_18_707}.
Namely, $\tau_{\text{DOS}}$, is given as
\begin{align}
	\tau_{\text{DOS}} &\sim \sum_{\sigma = \uparrow, \downarrow} D_{\text{L}, \sigma}(E) D_{\text{R}, \sigma}(E).
	\label{eq:dos_product}
\end{align}
Here, $D_{\text{L/R}, \sigma}(E)$ is the density of states of the left/right electrodes.
\par
In a similar manner to the Julliere's picture, 
we can consider the product of the LDOS inside the insulating barrier, $d_{\text{L/R}, \sigma}(E)$,
\begin{align}
	\tau_{\text{LDOS}} &\sim \sum_{i = 1}^{n_{\text{a}}}\sum_{\sigma = \uparrow, \downarrow} d_{\text{L}, i, \sigma}(E) d_{\text{R}, i, \sigma}(E),
	\label{eq:ldos_product}
\end{align}
where $d_{\text{L/R}, i, \sigma}(E)$ is the LDOS inside the barrier,
and $n_{\text{a}}$ is the number of atoms in the barrier layers which we focus on; 
layer-L and R.
By using the LDOS inside the barrier instead of the DOS of the electrodes,
we can take account of the details of the characters of materials used for the electrodes and barriers and also the decay inside the tunneling barrier,
which is in sharp contrast to $\tau_{\text{DOS}}$ where only the bulk properties of the metals are considered.
\par
To obtain the LDOS inside the barrier, we perform the non-scf calculation for the scattering region used in the calculation of the transmission, 
namely, $\mathrm{Ru}_{1-x}\mathrm{Cr}_{x}\mathrm{O}_{2}~(4~\mathrm{MLs})/\mathrm{TiO_{2}}~(9~\mathrm{MLs})/\mathrm{Ru}_{1-x}\mathrm{Cr}_{x}\mathrm{O}_{2}~(5~\mathrm{MLs})$, 
with the $\boldsymbol{k}$-point mesh of $21 \times 21 \times 1$ following the scf calculation. 
Then, we calculate the projected DOS onto each atom.
Here, we use the LDOS of the atoms in one layer away from the center of the barrier and take the product of the LDOS of the two atoms with the same $xy$-coordinates.
\section{Results and discussions}
\label{sec:results}
\subsection{Bulk property of $\mathrm{Cr}$-doped $\mathrm{RuO_{2}}$}
\label{subsec:bulk}
\begin{figure}[tbh]
	\centering
	\includegraphics[width=86mm]{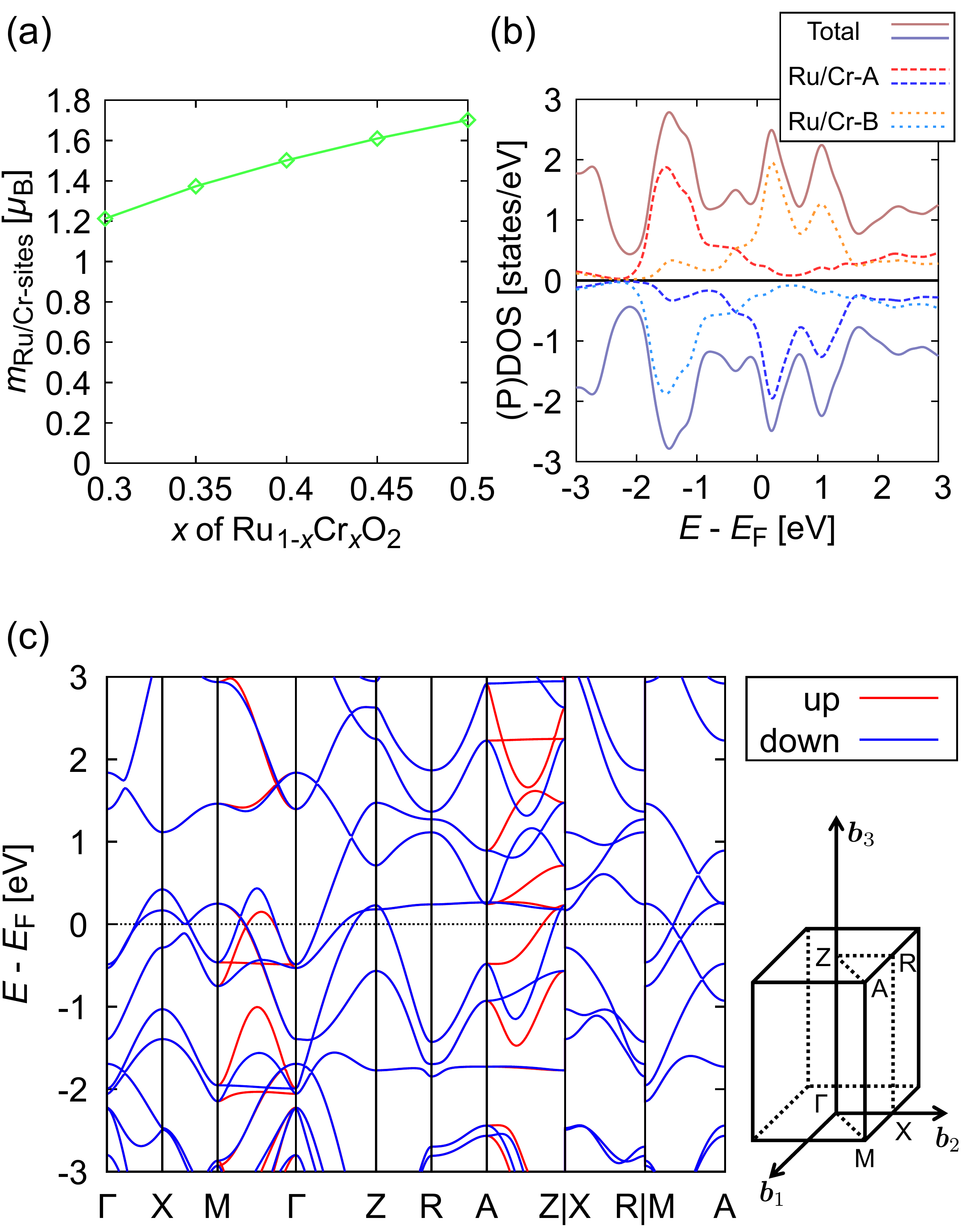}
	\caption{%
		(a) Averaged size of the magnetic moments of Ru(Cr) ions in $\mathrm{Ru}_{1-x}\mathrm{Cr}_{x}\mathrm{O}_{2}$ with respect to the amount of Cr-doping.
		(b) Spin-resolved density of states (DOS) and projected DOS (PDOS) of two Ru/Cr-sites for $\mathrm{Ru_{0.65}Cr_{0.35}O_{2}}$ as a function of energy.
		Positive (negative) values of the (P)DOS are the spin-up (down) components.
		(c) Energy band structure of $\mathrm{Ru_{0.65}Cr_{0.35}O_{2}}$ resolved by the spin degrees of freedom.
				The high-symmetry points in the Brillouin zone for the tetragonal crystal structure are schematically shown on the right side of the energy band,
				where $\boldsymbol{b}_{i}$ ($i = 1, 2, 3$) is the primitive reciprocal lattice vector.
	}
	\label{fig:rucro2_bulk}
\end{figure}
Before discussing the transmission properties of the MTJ with Cr-doped $\mathrm{RuO_{2}}$,
we discuss the bulk properties of $\mathrm{Ru}_{1-x}\mathrm{Cr}_{x}\mathrm{O}_{2}$.
We consider the systems with $0.3 \leq x \leq 0.5$ to ensure a magnetic moment large enough.
In this region, the collinear antiferromagnetic state has a lower energy than the nonmagnetic state.
We also confirm that the ferromagnetic state has a higher energy than the antiferromagnetic state for $0.3 \leq x \leq 0.5$ to exclude the possibility of realizing the ferromagnetic state because the rutile-type $\mathrm{CrO_{2}}$ has the ferromagnetic state~\cite{Chamberland1977_CritRevSolidStateMaterSci_7_1,Schwarz1986_JPhysFMetPhys_16_L211,Korotin1998_PhysRevLett_80_4305,Shim2007_PhysRevLett_99_057209}.
\par
In Fig.~\ref{fig:rucro2_bulk}(a), we show the Cr concentration dependence of the magnitude of the magnetic moments in $\mathrm{Ru}_{1-x}\mathrm{Cr}_{x}\mathrm{O}_{2}$.
The size of the magnetic moments becomes larger as the amount of Cr doping increases, 
which is consistent with the results of the previous study showing that the magnetic moments of Ru-ion increase by substituting the Ru-ion with Cr-ion~\cite{Wang2023_NatCommun_14_8240}.
This enhancement of the magnetic moment indicates that the Cr-doping strengthens the electronic correlation.
It should be noted that the Coulomb $U$ is not required to stabilize the magnetically ordered states here.
To discuss $\mathrm{RuO_{2}}$ as an antiferromagnet,
calculations with additional $U$ have been performed.
Meanwhile, in Ref.~\cite{Smolyanyuk2023_arXiv_2310.06909},
it has been pointed out that the actual Ru materials have a relatively small $U$, 
with which the stoichiometric $\mathrm{RuO_{2}}$ will not be an antiferromagnet.
Also, the observed values of the magnetic moment of the Ru-ions are $\sim 0.05$~$\mu_{\mathrm{B}}$ (0.23~$\mu_{\mathrm{B}}$) in the polarized (unpolarized) neutron diffraction measurement~\cite{Berlijn2017_PhysRevLett_118_077201},
which indicates that the electron correlation in $\mathrm{RuO_{2}}$ is not likely to be large.
In our case, the realization of the magnetic state is attributed not to the effect of $+U$ but to the replacement of Ru-$4d$ electrons with Cr-$3d$ electrons.
Hence, we expect that we naturally describe the Ru-based materials here.
\par
We show the total DOS and projected DOS (PDOS) of two Ru/Cr-sites resolved by the spin degrees of freedom for the $x = 0.35$ system in Fig.~\ref{fig:rucro2_bulk}(b).
The total DOS is symmetric with respect to the spin degrees of freedom,
which suggests that the magnetization in $\mathrm{Ru}_{1-x}\mathrm{Cr}_{x}\mathrm{O}_{2}$ is compensated in total.
We also find that each of the two Ru/Cr-sites in a unit cell has the spin-polarized PDOS,
which is consistent with finite magnetic moments of Ru-sites.
\par
Figure~\ref{fig:rucro2_bulk}(c) shows the energy band structure of $\mathrm{Ru}_{1-x}\mathrm{Cr}_{x}\mathrm{O}_{2}$.
We see the spin splitting along $\mathrm{M}$--$\mathrm{\Gamma}$ and $\mathrm{A}$--$\mathrm{Z}$ lines in the momentum space,
where $|k_{x}| = |k_{y}|$.
which is the same as the spin splitting band observed in the pure $\mathrm{RuO_{2}}$~\cite{Ahn2019_PhysRevB_99_184432,Smejkal2020_SciAdv_6_eaaz8809}.
\subsection{Tunnel magnetoresistance effect with $\mathrm{Cr}$-doped $\mathrm{RuO_{2}}$ electrode}
\label{subsec:tmr}
\begin{figure}[tbh]
	\centering
	\includegraphics[width=86mm]{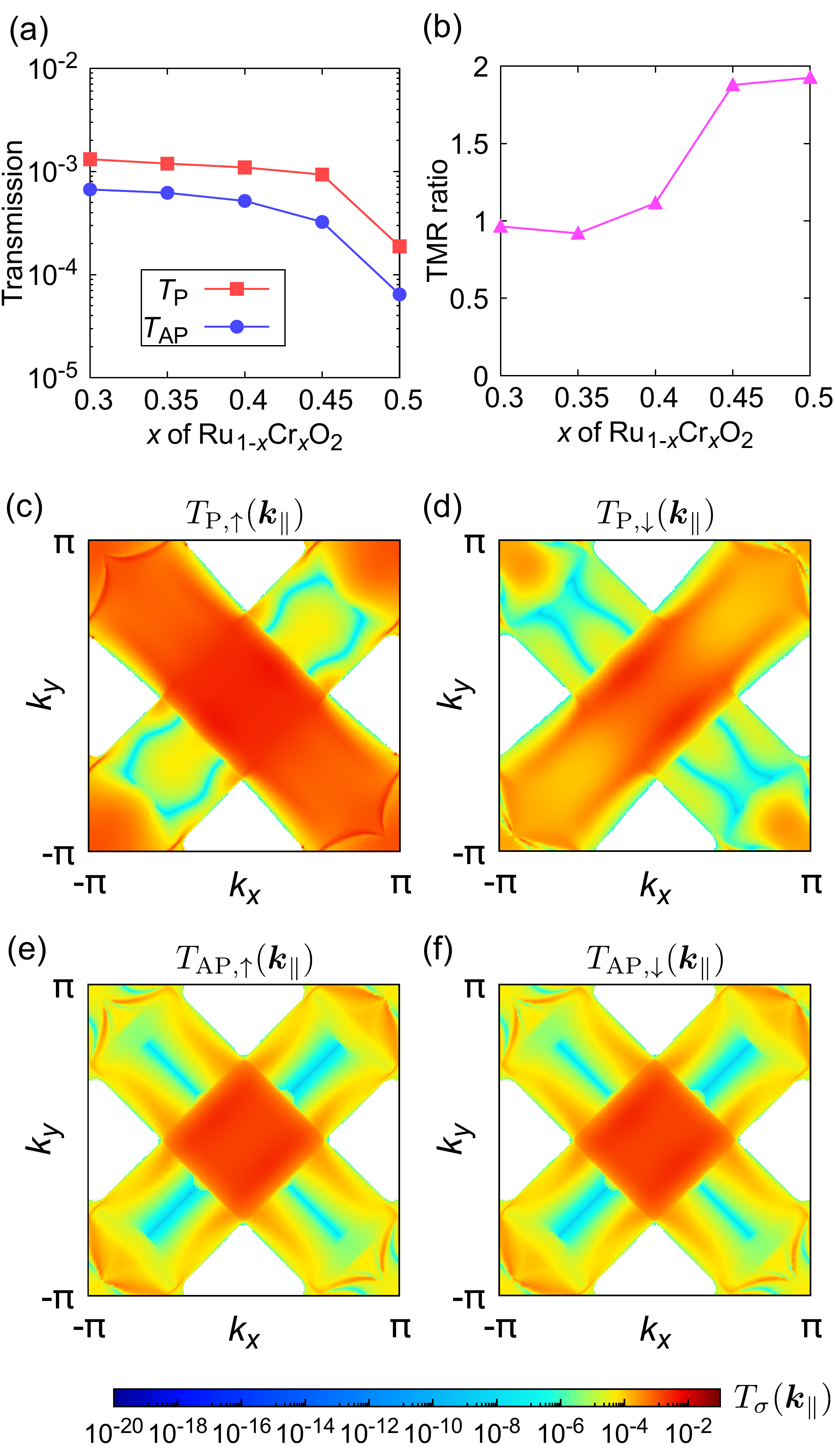}
	\caption{%
		(a) Cr-doping amount dependence of the total transmission for the parallel and antiparallel configurations of the $\mathrm{Ru}_{1-x}\mathrm{Cr}_{x}\mathrm{O}_{2}/\mathrm{TiO_{2}}/\mathrm{Ru}_{1-x}\mathrm{Cr}_{x}\mathrm{O}_{2}$ magnetic tunnel junction.
		(b) TMR ratio calculated by $\left( T_{\text{P}} - T_{\text{AP}} \right) / T_{\text{AP}}$.
		(c)--(f) Transmission resolved by the spin and the in-plane momentum $\boldsymbol{k}_{\parallel}$, $T_{\sigma}(\boldsymbol{k}_{\parallel})$, 
		for the $x = 0.35$ system.
		Transmission of (c) majority and (d) minority spins for the parallel configuration.
		Transmission of (e) majority and (f) minority spins for the antiparallel configuration.
	}
	\label{fig:transmission}
\end{figure}
Next, we discuss the TMR effect in the $\mathrm{Ru}_{1-x}\mathrm{Cr}_{x}\mathrm{O}_{2}/\mathrm{TiO_{2}}/\mathrm{Ru}_{1-x}\mathrm{Cr}_{x}\mathrm{O}_{2}$ MTJ.
In Fig.~\ref{fig:transmission}(a), we show the Cr concentration dependence of the total transmission $T_{\text{tot}}$ of the parallel and antiparallel configurations at the Fermi level, $T_{\text{P}}$ and $T_{\text{AP}}$, respectively.
Both $T_{\text{P}}$ and $T_{\text{AP}}$ decrease as the amount of the Cr-doping increases.
In $0.3 \leq x \leq 0.5$, $T_{\text{P}}$ is larger than $T_{\text{AP}}$.
This means that the corresponding TMR ratio, 
defined by $\left( T_{\text{P}} - T_{\text{AP}} \right) / T_{\text{AP}}$,
takes positive finite values as shown in Figure~\ref{fig:transmission}(b).
The TMR ratio is around 100\%--200\%.
This value is smaller than that in the $\mathrm{RuO_{2}}/\mathrm{TiO_{2}}/\mathrm{RuO_{2}}$ MTJ~\cite{Shao2021_NatCommun_12_7061}.
Still, we should note that a large $U$ is often assumed for $\mathrm{RuO_{2}}$,
which possibly overestimate the magnetism of pure $\mathrm{RuO_{2}}$ as discussed in Sec.~\ref{subsec:bulk}. 
In addition, we do not add the Coulomb $U$ here, 
which makes us underestimate the energy gap of $\mathrm{TiO_{2}}$ here.
If we take the effect of $+U$ into account and evaluate the band gap of $\mathrm{TiO_{2}}$ more accurately,
the TMR ratio in the $\mathrm{Ru}_{1-x}\mathrm{Cr}_{x}\mathrm{O}_{2}$-based MTJ may become as large as that in the $\mathrm{RuO_{2}}$-based MTJ, 
which we leave as the future work.
\par
In Fig.~\ref{fig:transmission}(c)--\ref{fig:transmission}(f),
we show the transmission at the Fermi level resolved by the in-plane momentum perpendicular to the conducting path $\boldsymbol{k}_{\parallel}$.
For the parallel configuration, the transmission takes a large value near the $|k_{x}| = |k_{y}|$ line,
where the spin splitting of $\mathrm{Ru}_{1-x}\mathrm{Cr}_{x}\mathrm{O}_{2}$ is present as seen in the band structure shown in Fig.~\ref{fig:rucro2_bulk}(b).
For the antiparallel configuration, since the two magnetic electrodes have the opposite spin polarization with each other,
$T(\boldsymbol{k}_{\parallel})$ around the $\Gamma$-point takes a large value.
\subsection{Local density of states inside the barrier}
\begin{figure}[tbh]
	\centering
	\includegraphics[width = 86mm]{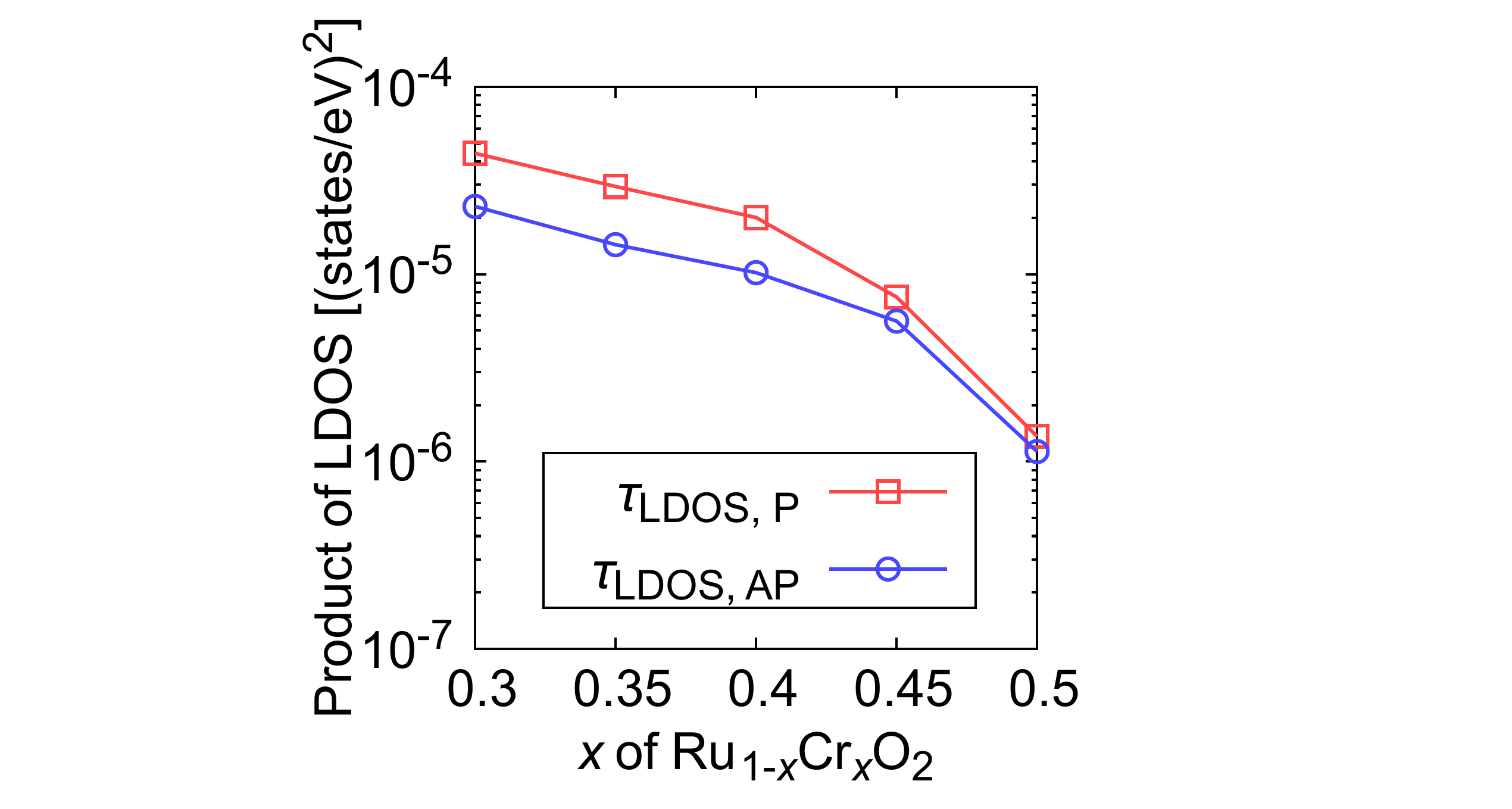}
	\caption{%
		Cr-concentration dependence of the product of the local density of states at the Fermi energy for parallel and antiparallel configurations calculated by Eq.~(\ref{eq:ldos_product}).
		\label{fig:ldos}
	}
\end{figure}
Finally, we discuss the correspondence between the transmission and the LDOS inside the barrier.
Figure~\ref{fig:ldos} shows the product of the LDOS at the Fermi energy calculated following  Eq.~(\ref{eq:ldos_product}) with respect to the amount of Cr.
We find that the product of the LDOS decreases for each of the parallel and antiparallel configurations, 
and $\tau_{\text{LDOS, P}}$ takes larger values than $\tau_{\text{LDOS, AP}}$.
These features are consistent with the transmission property shown in Fig.~\ref{fig:transmission}(b),
which indicates that we can qualitatively capture the transmission behavior using the LDOS not only in the idealized lattice models~\cite{Tanaka2023_PhysRevB_107_214442} but also in a realistic system.
\section{Summary}
In summary, we have discussed the tunnel magnetoresistance (TMR) effect using the Cr-doped $\mathrm{RuO_{2}}$ as an electrode of the magnetic tunnel junction (MTJ) from first-principles.
We have performed \textit{ab initio} calculation of the tunneling conductance in the $\mathrm{Ru}_{1-x}\mathrm{Cr}_{x}\mathrm{O}_{2}/\mathrm{TiO_{2}}/\mathrm{Ru}_{1-x}\mathrm{Cr}_{x}\mathrm{O}_{2}$ MTJ.
We have found that a finite TMR effect is generated by a finite spin splitting in the momentum space,
which is supported by the enhancement of the electron correlation in $\mathrm{Ru}_{1-x}\mathrm{Cr}_{x}\mathrm{O}_{2}$ owing to the Cr-doping.
We have also shown that the qualitative nature of the obtained TMR effect can be traced using the local density of states (LDOS) inside the barrier of the MTJ.
We believe that this correspondence between the TMR property and the LDOS in realistic systems will be helpful in searching for materials suitable for MTJs.
\begin{acknowledgments}
K.T. is grateful to Meng Wang, Fumitaka Kagawa, and Shiro Sakai for the collaboration on Ref.~\cite{Wang2023_NatCommun_14_8240},
which motivates us to work on this subject.
K.T. also thanks Takashi Koretsune and Yuta Toga for useful comments.
This work was supported by JST-Mirai Program (Grant No.~JPMJMI20A1), JST-CREST (Grant No.~JPMJCR23O4), JST-ASPIRE (Grant No.~JPMJAP2317), 
and JSPS-KAKENHI (Grant No.~21H04437, No.~JP21H04990, No.~JP22H00290, No.~JP24K00581).
We use the \textsc{VESTA} software~\cite{Momma2011_JApplCryst_44_1272} to visualize the crystal structure with the aid of \textsc{XCrySDen} software~\cite{Kokalj1999_JMolGraphicsModelling_17_176} to generate an input file.
\end{acknowledgments}
%

\end{document}